\documentclass[aps,pra,amsmath,notitlepage,nofootinbib,twocolumn,superscriptaddress,noeprint]{revtex4-2}
\usepackage[utf8]{inputenc}
\usepackage[english]{babel}
\usepackage[T1]{fontenc}
\usepackage{amsmath,amsfonts,amssymb,amsthm,bm,bbm,bbold}
\usepackage{graphicx}
\graphicspath{{Figures/}{./}}
\usepackage{amssymb}
\usepackage{mathtools}
\usepackage{physics}
\usepackage{makecell}
\usepackage{textcomp}
\usepackage{microtype}
\usepackage[dvipsnames]{xcolor}
\usepackage{dsfont}
\usepackage[breaklinks=true,colorlinks=true,linkcolor=red,urlcolor=teal,citecolor=red]{hyperref}
\usepackage[caption=false]{subfig}

\newcommand{\be}{\begin{equation}}
\newcommand{\ee}{\end{equation}}
\newcommand{\ben}{\begin{eqnarray}}
\newcommand{\een}{\end{eqnarray}}
\newcommand{\bes}{\begin{subequations}}
\newcommand{\ees}{\end{subequations}}
\newcommand{\bF}{\begin{figure}}
\newcommand{\eF}{\end{figure}}

\def\ket#1{ | #1 \rangle}

\newcommand{\avg}[1]{\langle #1 \rangle}
\newcommand{\proj}[1]{| #1 \rangle \langle #1 |}

\let\originalleft\left
\let\originalright\right
\renewcommand{\left}{\mathopen{}\mathclose\bgroup\originalleft}
\renewcommand{\right}{\aftergroup\egroup\originalright}
\renewcommand{\right}{\aftergroup\egroup\originalright}

\usepackage{stackengine}

\begin{document}

\title{Sensing with Quantum Light: A perspective}


\author{Animesh Datta}
\email{animesh.datta@warwick.ac.uk}
\affiliation{Department of Physics, University of Warwick, Coventry, CV4 7AL, United Kingdom}


\date{\today}

\begin{abstract}
I present my perspective on sensing with quantum light. 
I summarise the motivations and methodology for identifying quantum enhancements in sensing over a classical sensor.
In the real world, this enhancement will be a constant factor, and not increase with the size of the quantum probe as 
is often advertised. 
I use a limited survey of interferometry, microscopy, and spectroscopy to extract the vital 
challenges that must be faced to realise tangible enhancements in sensing with quantum light.
\end{abstract}

\maketitle

\section{Introduction}

Over the ages, light has been central to sensing and detecting phenomena in the Natural world across length and timescales, from observational cosmology to nanoscopy. 
Light also happens to be the medium whose quantum properties are most readily redolent in ambient conditions~\cite{Walmsley2015}.
Thus, it is only natural that sensing with quantum light has been investigated~\cite{Caves1981} and pursued with some vigour over the last decade~\cite{Schnabel2010,Taylor2016,Lawrie2019,Berchera2019,Szoke2020,Lee2021,Kim2023,Defienne2024,Moreva2025}.
It has enabled us to see things that would have been impossible without it~\cite{Ganapathy2023}.

My endeavour in this perspective on sensing with quantum light is to note some past advances and future challenges.
Rather than a review, I aim to identify the underlying commonalities -- in existing methodology and foreseeable problems.
My choice of material is evidently selective. 
I focus on the three sensing modalities of interferometry, microscopy, and spectroscopy. 
I choose them because they form large classes of sensing applications classically and have the potential of benefitting from quantum light.
They also encapsulate amongst themselves the vital aspects of the principle and practice of sensing with quantum light and are naturally amenable to nanophotonics.

\begin{figure}[t]
\includegraphics[width=9cm]{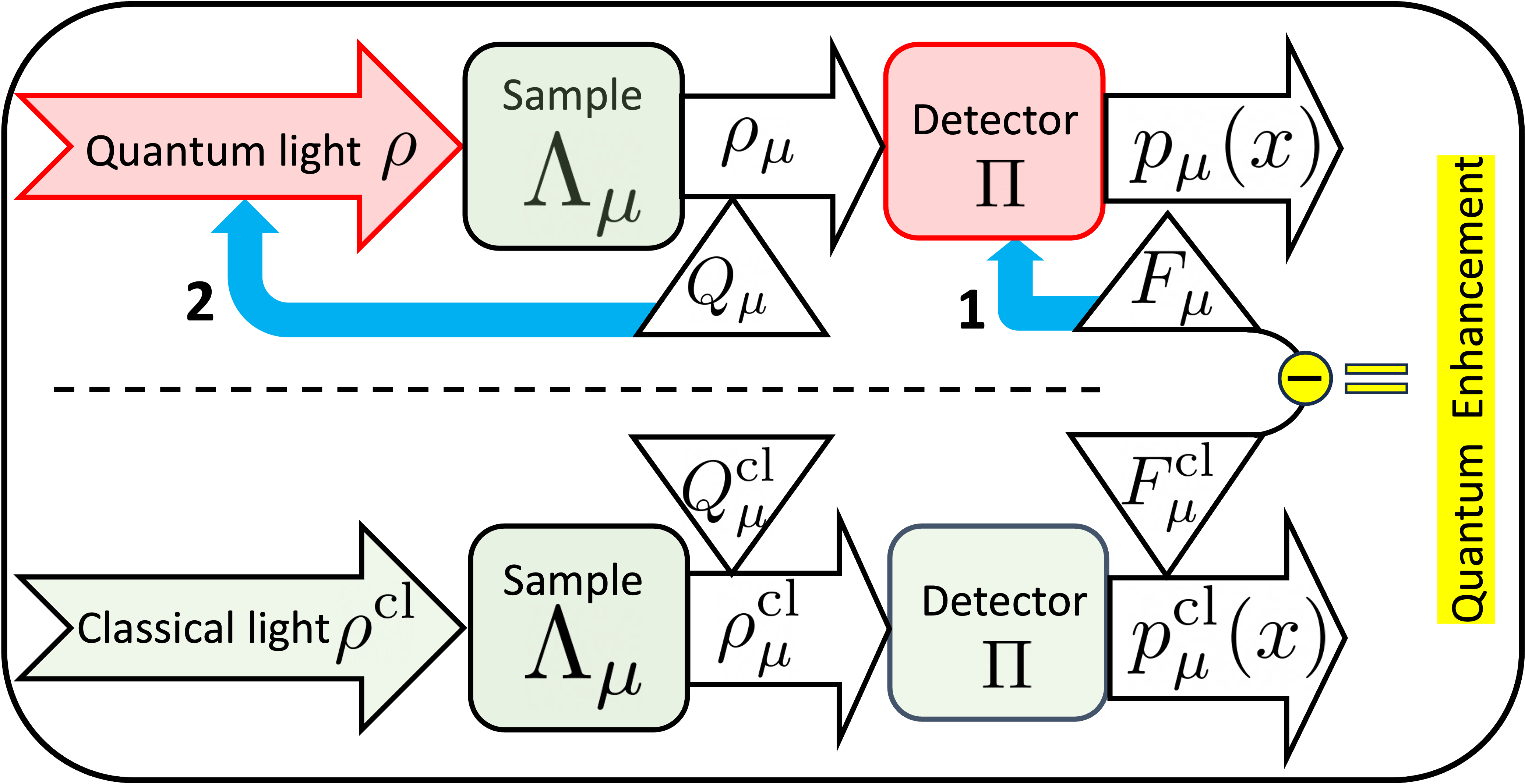}
\caption{The layout of sensing with quantum light (above dashed line). Corresponding sensing task with classical light (below dashed line).
The conceptual route to identifying tangible quantum enhancement lies in the difference of the two classical Fisher informations $F_{\mu}$ and $F_{\mu}^{\mathrm{cl}}.$
This quantum enhancement can only be established experimentally. 
The symbols are defined in Sec.~\ref{sec:sensing} and $\rho^{\mathrm{cl}}$ is described in Sec.~\ref{sec:light}.
The two blue arrows denote the two optimisations in Sec.~\ref{sec:qest}, of the two red boxes. 
The green boxes are typically fixed in advance.}
\label{fig:layout}
\end{figure}

I begin with the mathematical formalism that captures sensing in Sec.~\ref{sec:sensing}.
It presents the classical and quantum Fisher information as quantities to be evaluated to begin the process of identifying a quantum enhancements in sensing.
I summarise its conceptual message -- of identifying quantum enhancement due to quantum light over classical light in Fig.~\ref{fig:layout}.
Establishing tangible enhancement of sensing quantum light can only be done experimentally.
In Sec.~\ref{sec:light}, I encapsulate our understanding of classical and quantum light.

I discuss the illusory quadratic quantum scaling that is often the objective of quantum sensing studies in Sec.~\ref{sec:elus}.
I emphasise why it is impossible in the real world, and note how tangible quantum advantages may actually be attained.
In Sec.~\ref{sec:mod}, I present some advances in interferometry, microscopy, and spectroscopy with quantum light. 
Rather than an exhaustive record, I select works to identify the main challenges in sensing with quantum light.
I end in Sec.~\ref{sec:Conc} with a collation of these challenges and avenues to directing efforts to overcome them.

\section{Sensing} 
\label{sec:sensing}

The mathematical formalism describing sensing is a part of information theory and statistics~\cite[Chapter 11]{ThomasCover2006}.
It is formally called estimation theory. 
In the following, I note the classical aspects of the formalism essential for a swift transition to the quantum.
While nothing in this formalism is particular to sensing with light and applies in the abstract to any sensor, I will resort to the former for concreteness.

A sensor operates by estimating certain parameters of the light field at its detector.
It matters not whether the light came from a distant star or an adjacent nanostructure. 
The recorded data at the detector $\{x_1,x_2,\cdots,x_r \}$ is described mathematically as a statistical model with probability measure $p_{\mu}(x) \equiv p(x|\mu),$ where $\mu$ 
denotes the parameter\footnote{I focus on a single parameter. The classical estimation of multiple parameters simultaneously in not conceptually different. The quantum estimation of multiple parameters simultaneously is a fascinating area due to the non-commutativity of measurements. I do not address this here. See, for instance, Ref.~\cite{Albarelli2020}.} to be estimated and $x$ labels the output of the detector, such as current, wavelength or photon number.
$r$ is the number of recorded data points or repetitions.
I choose $x$ and $\mu$ to be continuous for now.
The performance of a sensor is captured by the accuracy  and error of the estimate it provides.
If this estimate is denoted by $\tilde\mu(x),$ its accuracy is proportional to 
$\tilde\mu(x) - \mu$ and error to $(\tilde\mu(x) - \mu)^2.$
The smaller these quantities, the better the sensor.
 
 As the data recorded at the detector is statistical, the performance of a sensor are formally captured by its bias and precision.
The former is the expectation value of the accuracy, defined as 
$b(\tilde \mu) = \mathbb{E} \left[  \tilde\mu(x) - \mu \right] = \int p_{\mu}(x) \tilde\mu(x) ~\mathrm{d}x - \mu,$
where I have used $ \int p_{\mu}(x) ~\mathrm{d}x =1$ for a probability measure.
The latter is the expectation value of the error, defined as the variance
\be
\label{eq:var}
\mathrm{Var}_{\mu} = \mathbb{E} \left[  (\tilde\mu(x) - \mu)^2 \right] = \int p_{\mu}(x) (\tilde\mu(x) - \mu)^2 ~\mathrm{d}x.
\ee
$\mathrm{Var}_{\mu}$ is often called the mean square error. Its square root is the standard deviation.

\subsection{Classical estimation theory}

To deliver the goal of designing a better sensor, estimation theory is used to identify the smallest possible value of $\mathrm{Var}_{\mu}$ allowed by laws of probability theory.
A central result in classical estimation theory, known as the Cram\'er-Rao bound states~\cite{ThomasCover2006}
\be
r \mathrm{Var}_{\mu} \geq \frac{\left(1 + b'(\tilde \mu)\right)^2}{F_{\mu}} + b(\tilde \mu)^2,
\ee
 where 
 \be
 \label{eq:fi}
 F_{\mu}  = \mathbb{E} \left[ ( \partial_{\mu} \ln p_{\mu}(x) )^2 \right] =  \int \frac{(\partial_{\mu} p_{\mu}(x) )^2}{p_{\mu}(x)}  ~\mathrm{d}x
 \ee
 is the classical Fisher information (CFI).  
 For an unbiased estimator, $b(\tilde \mu) = 0$ and  
 \be
 \label{eq:crbb}
\mathrm{Var}_{\mu} \geq \frac{1}{r F_{\mu}} .
\ee
This inequality is saturated with unbiased estimators under certain regularity conditions on $p_{\mu}(x).$
For instance, the maximum likelihood estimator achieves the equality asymptotically~\cite{ThomasCover2006}.

The Cram\'er-Rao bound shows that variance of an estimator cannot be be made arbitrarily small.
It thus places a mathematical limit on how precise any sensor can be.
This limit is imposed by the amount of information that can extracted about a parameter from recorded data.
More data or larger $r$ leads to a lower variance.
From Eq.~\eqref{eq:fi}, it follows that a larger Fisher information and consequently lower variance relies on the recorded data $p_{\mu}(x)$ varying strongly with the parameter $\mu.$
This is to be expected and is intuitively familiar to all sensor designers.

\subsection{Quantum estimation theory}
\label{sec:qest}


Born rule is the ingredient of quantum mechanics that pertains most directly to sensing.
This is unsurprising, as it connects quantum states with observable probabilities.
This is the starting point of quantum estimation theory, pioneered in the works of  Helstrom~\cite{Helstrom1976} and Holevo~\cite{Holevo2011} in quantum information theory.

Mathematically,
\be
\label{eq:born}
p_{\mu}(x) = \Tr[ \rho_{\mu} \Pi_x],
\ee
where $\rho_{\mu}$ denotes the quantum state of the light field which depends on the parameter $\mu$ to be estimated and $\Pi_x$ is the operator describing the detector
outcome $x$, formally called the positive operator-valued measure (POVM) $\bm{\Pi} = \{ \Pi_x\}$~\cite{Nielsen2010}.
Inserting Eq.~\eqref{eq:born} in Eq.~\eqref{eq:var} shows that the variance $\mathrm{Var}_{\mu}$ depends on the quantum state $\rho_{\mu}$ and the choice of the detector $\bm{\Pi}.$
The same holds for the CFI.

This sets the stage for quantum-enhanced sensing, which is a two-step process to improve the precision of a sensor, or reduce the corresponding variance by choosing $\bm{\Pi}$ and 
$\rho_{\mu}$ judiciously.
See Fig.~\eqref{fig:layout}.

The first step is to minimise the variance $\mathrm{Var}_{\mu}$ over all physically allowed detectors, that is, POVMs $\bm{\Pi}.$
Combining this with Eq.~\eqref{eq:crbb} results in the quantum Cram\'er-Rao bound
 \be
 \label{eq:qcrb}
\mathrm{Var}_{\mu} \geq \frac{1}{r F_{\mu}} \geq \frac{1}{r Q_{\mu}},
\ee
where $Q_{\mu} = \max_{\bm{\Pi}} F_{\mu}$ is the quantum Fisher information (QFI). 
The minimisation of the variance translates into a maximisation of the CFI due to reciprocal relation in Eq.~\eqref{eq:crbb}.
This is denoted by the blue arrow numbered \textbf{1} in Fig.~\eqref{fig:layout}.
Mathematically~\cite{Helstrom1976}
 \be
 \label{eq:sld}
Q_{\mu}  = \Tr \left[\rho_{\mu}L^2 \right], \text{where~}\rho_{\mu}L + L\rho_{\mu} = 2 \partial_{\mu}\rho_{\mu}.
 \ee
The hermitian operator $L$  is known as the symmetric logarithmic derivative.
The eigenvectors of $L$ are the POVM elements $\Pi_x$ that minimise the variance.

As in classical estimation, the amount of information that can be extracted from a quantum state depends on how strongly it changes with the parameter $\mu.$
The operator $L$ is proportional to this rate of change, with the symmetrisation necessary to account for the non-commutative nature of the Hilbert space in which $\rho_{\mu}$ resides.
The precision of a quantum sensor thus depends solely on the quantum state $\rho_{\mu}$ and its derivative with respect to the parameter 
$\mu.$\footnote{
I focus solely on the scenario where the parameter $\mu$ has a unique true value. Another typical scenario is when the parameter takes values from a 
prior distribution $P_{\mathrm{pri}}(\mu).$ Then the minimum mean square error (MMSE) is given by~\cite{Personick1971}
\be
\text{MMSE} = \min_{M} \int P_{\mathrm{pri}}(\mu) \Tr \left[\rho_{\mu} (M - \mu I)^2 \right] ~\mathrm{d} \mu,
\ee
where the minimisation is undertaken over all hermitian operators $M.$
The operator $M_*$ attaining this minimum is given by the solution of~\cite{Personick1971}
\be
\rho^{(1)}M_* + M_* \rho^{(1)} = 2 \rho^{(2)},
\ee
where $\rho^{(1)} = \int P_{\mathrm{pri}}(\mu) \rho_{\mu} \mathrm{d} \mu, \rho^{(2)} = \int \mu P_{\mathrm{pri}}(\mu) \rho_{\mu} \mathrm{d} \mu.$
Note the similarities to and differences from the structure of Eq.~\eqref{eq:sld}.
}

Quantum sensors are, at least in principle, no worse than classical sensors.
This is because the QFI is greater than or equal to the CFI, that is 
$Q_{\mu} = \max_{\bm{\Pi}} F_{\mu} \geq F_{\mu}.$ 
Making this relevant in practice is the fact that there always exists an optimal POVM $\bm{\Pi}_{*}$  for which the CFI equals the QFI, that is, 
the second inequality in Eq.~\eqref{eq:qcrb} can be saturated.
The ease with which $\bm{\Pi}_{*}$ may be implemented physically is a technological issue we address later.

The second step is to maximise the QFI $Q_{\mu}$ over probe quantum states $\rho,$ where $\rho_{\mu} = \Lambda_{\mu}[\rho]$ denoted by the blue arrow numbered \textbf{2} in Fig.~\eqref{fig:layout}.
Here $\Lambda_{\mu}[\cdot]$ captures the physics of the sensor that imprints the parameter $\mu$ on $\rho.$ 
For a light-based sensor, this may range from a linear optical element such as a phase shift to an elaborate spectroscopic setup 
involving nonlinear interactions of a complex quantum system with one or more pulses of light.
Mathematically, these can all be described by a completely-positive, trace-preserving (CPTP) map~\cite{Nielsen2010} which can be obtained from the
 underlying physical model of the sensor capturing the interaction of the  parameter of interest $\mu$ with the quantum state $\rho$ of the probe light field.
The geometry of the space of CPTP maps plays a central role in determining the performance of quantum sensors in the real world. See Sec.~\ref{sec:elus}.

A quantum sensor using the probe $\rho$ is, in principle, better than its classical counterpart, if the QFI corresponding to $\rho_{\mu} ,$
exceeds that of $\rho_{\mu}^{\mathrm{cl}}  = \Lambda_{\mu}[\rho^{\mathrm{cl}}],$
where $ \rho^{\mathrm{cl}}$ corresponds to classical light. 
The latter QFI is denoted as $Q_{\mu}^{\mathrm{cl}}$ in Fig.~\eqref{fig:layout}.
In practice, a quantum sensor using the probe $\rho$ is better than its classical counterpart, if $F_{\mu} > F_{\mu}^{\mathrm{cl}},$ that is the former has a higher CFI,
as depicted in Fig.~\eqref{fig:layout}.
 
But what is ``classical'' light?

\section{Light} 
\label{sec:light}

``Classical'' light may, strictly speaking, be considered an oxymoron.
This is because light, more formally the electromagnetic field, has a fundamental quantum mechanical description due to Dirac.  
A very lucid exposition was provided by Fermi~\cite{Fermi1932}.
Like many orxymoronic expressions, however, ``classical'' light has a purchase due to reason, convenience and utility, as I summarise below.

\subsection{Classical light}

Following the successful development of quantum electrodynamics~\cite{Schweber1994}, attention reverted to the quantum mechanical description of optics -- quantum optics~\cite{Glauber1962,Sudarshan1963,Glauber1963}. It showed that light fields generated by arbitrary distributions of classical currents
 have an especially simple description in terms of coherent states~\cite{Glauber1962}. 
 The coherent state of the light field of a single mode is thus identified with ``classical'' light, as is the Poisson distribution of the average occupation number of the $n$th Fock state 
 $\ket{n}$ -- the quantum state with exactly $n$ photons.
 In this basis, the coherent state is expressed as 
 \be
 \ket{\alpha} = \exp(-\frac{|\alpha|^2}{2}) \sum_{n=0}^{\infty} \frac{\alpha^n}{\sqrt{n!}} \ket{n},
 \ee
where $\alpha$ is a complex number proportional to amplitude of electric field.
In my notation, $\rho^{\mathrm{cl}} = \proj{\alpha}.$

The coherent state $\rho^{\mathrm{cl}}$ is, in fact, the quantum mechanical ground state of the hamiltonian of the free quantum electromagnetic field in absence of charges and currents.
This hamiltonian corresponds to that of an isolated one-dimensional harmonic oscillator~\cite{Glauber1963}.
A coherent state is thus a minimum uncertainly wavepacket in the phase space spanned by the position $q$ and momentum $p$ of the oscillator, and satisfies
\be
\label{eq:zp}
(\Delta^2 \hat p) (\Delta^2 \hat q) = \frac{\hbar^2}{4}.
\ee
Here $ \Delta^2 \hat O = \avg{\hat{O}^2} - \avg{\hat O}^2$ is the variance and $\avg{\cdot}$ denotes the expectation value of $\hat O,$ 
in this instance in the quantum state $\rho^{\mathrm{cl}}.$
Conceptually, the coherent state thus represents as close an approach to classical localisation as possible~\cite{Glauber1963}.

A coherent state is also very close to the output state of a laser near its threshold~\cite{GeaBanacloche1998}.
The identification of coherent states as classical light is thus both conceptual and operational, the latter stemming from the ubiquity of high quality of lasers for swathes of the electromagnetic spectrum.
I should emphasise that both these routes, via Eq.~\eqref{eq:zp} and the laser, are fundamentally quantum mechanical.

\subsection{Quantum light}
\label{sec:qlight}

In the parlance, nonclassical or quantum light is a light field whose state does not have a simple description in terms of coherent states.
Technically,  a state of light is nonclassical if its Glauber–Sudarshan $P$-function is not a positive probability density function.
Resting on a negative, this definition is quantitively unwieldy. Consequently, a plethora of measures of nonclassicality have developed over time~\cite{Chuan2019},
including those relying on other distribution functions~\cite{Hillery1984}.
All these measures seek to capture how different a given state of the light field is from a coherent state.

Generating quantum light involves acting on the output of a laser in some way, typically via a nonlinear material.
While there is an infinitude of possible quantum states of light, much effort has been directed to generating states with exactly one~\cite{Meyer-Scott2020}.
and two photons~\cite{Solntsev2017}. 
The former often rely on the saturation nonlinearity of a quantum emitter. 
The emitted single photons can, in principle, be interfered in quantum networks of linear optical elements to produce quantum states of light with a larger number of photons~\cite{Clements2016}.

Quantum states of light can also have an indefinite number of photons. The most prominent in this class are squeezed states~\cite{Schnabel2017}. 
These are, like coherent states, minimum uncertainly wavepackets satisfying Eq.~\eqref{eq:zp}, but with the variance in one of the quadratures reduced 
(or squeezed) at the expense of the other.
Their generation relies in the second or third order nonlinear susceptibility of a material.
Other quantum states of light can be produced by interfering definite and indefinite photon number states~\cite{Bartley2012}.
Yet others, such as superpositions of coherent states called Schr\"odinger cat states, can be obtained by subtracting photons from squeezed states~\cite{Wakui2007,Bartley2013}.

\subsection{Modes, entanglement, and confusion}

The quantum states of light I noted above reside in a single mode.
A mode is a solution of the electromagnetic wave equation given a set of boundary conditions~\cite{Dandliker1999}.
They play a central role in the design and operation of nanophotonic devices~\cite{Lu2013}.
Quantum states of light can also be constructed that reside across multiple modes.
They could also contain definite or indefinite numbers of photons. An example of the latter is a multimode squeezed state~\cite{Cai2017}.

Quantum entanglement is another notion of nonclassicality.
An entangled quantum state of two or more subsystems is defined as one that is not separable, that is, it cannot be expressed 
a mixture of tensor products of quantum states belonging to the individual subsystems~\cite{Nielsen2010}.
This is distinct from a state not having a simple description in terms of coherent states, as in Sec.~\ref{sec:qlight}.
An example is a single-mode squeezed state. It is a nonclassical state but not an entangled state.
Quantum light may have entanglement between photons residing in different modes~\cite{WALBORN2010}. 

The indistinguishability of photons in the same mode sometimes confuses matters. 
This is because the identification of distinct (or distinguishable) subsystems of indistinguishable particles (photons) is not possible.
Thus, entanglement between photons in the same mode may be judged unphysical.
Indeed, efforts to imbue it with operational information theoretic meaning resorts to the use of modes~\cite{BENATTI2020}

\section{Quantum sensing -- elusive scaling}
\label{sec:elus}

As the electromagnetic field is defined by an amplitude and a phase, the categories of parameters it can sense are losses, relative phases, and their combinations.
Amongst these, estimating the difference in the path lengths of light travelling by two different routes from a source to a detector is perhaps the most elementary. 
This can be cast as relative phase sensing, and captures applications ranging from gravitational wave detection~\cite{Caves1981} 
to phase-contrast imaging~\cite{Humphreys2013}.
The formal study of quantum sensing with light has its origins in the 
work of Helstrom~\cite{Helstrom1976} and later in the
development of laser-interferometric gravitational wave detectors\footnote{\url{https://www.ligo.caltech.edu/system/media_files/binaries/386/original/LIGOHistory.pdf}}, which are in effect Michelson interferometers.

Using the formalism from Sec.~\ref{sec:qest} with $\mu \equiv \phi$ a relative phase parameter,
 $\Lambda_{\mu} \equiv U_{\phi} = \exp{-i \phi \hat{n}}$ is a unitary map where $\hat{n}$ is the number operator. Then,
$\rho_{\phi} = U_{\phi} \rho U^{\dag}_{\phi},$ and the QFI $Q_{\mu} = \Delta^2 \hat n,$ now evaluated for input state $\rho$ into the interferometer~\cite{Braunstein1996}.
Maximising this, as per the second step in Sec.~\ref{sec:qest} over quantum states of light with $N$ photons -- exactly or on average~\cite{Lang2014}, gives
\be
\label{eq:sq}
Q_{\phi} \sim N^2.
\ee
This quadratic dependence of the QFI of estimating the relative phase on the number of photons is hailed as the hallmark of quantum-enhanced sensing,
compared to the QFI for classical light $Q_{\phi} \sim N$ where $N=|\alpha|^2$ for $\rho^{\mathrm{cl}} = \proj{\alpha}.$

The quadratic dependence in Eq.~\eqref{eq:sq} is often dubbed the Heisenberg limit or scaling~\cite{Holland1993}, though its connection to the uncertainly relation in Eq.~\eqref{eq:zp} is tenuous.
The quantum states that attain the quadratic scaling in the photon number (exact or average) are certainly nonclassical, 
and include instances such as N00N states and squeezed states~\cite{Lang2014}.
Aspiring for this quadratic scaling in phase estimation has driven much of the interest in quantum-enhanced sensing~\cite{Giovannetti2004,Giovannetti2011},
rooted in seeking a more precise sensor using fewer probe resources.
I emphasise that the number of photons actually used is $rN.$ The scaling in $r$ is entirely of classical and statistical origin and not subject to quadratic quantum enhancements.

A central challenge is the production of quantum states of light with large photon numbers at a repetition rate 
comparable to a coherent state which can easily have billions or trillions of photons.
`Bright' squeezed states with comparable mean photon numbers have been reported recently~\cite{Rasputnyi2024}.
Preparing quantum states with high fixed photon number remains hard despite novel theoretical ideas~\cite{Rivera2023,Heras2024}.
Should such ideas become realisable, their integration in to the wider quantum photonic sensing architecture would be most welcome.

The quadratic scaling  for phase estimation is impossible in real-world sensing scenarios even if desirable quantum states are available.
This is because all real world sensors inevitably encounter losses and other noise or decoherence processes~\cite{Demkowicz-Dobrzanski2012}. 
Consequently, the best possible scaling is 
\be
\label{eq:lin}
Q_{\phi} \sim cN,
\ee
where $c$ is a constant independent of $N.$ It depends on the nature and strength of the loss, noise or decoherence processes.

The quadratic scaling for phase estimation in Eq.~\eqref{eq:sq}
 in the absence of loss and noise also applies to any parameter $\mu$ generated by a $\Lambda_{\mu}$ that is a unitary map.
 The concomitant linear scaling in Eq.~\eqref{eq:lin} then applies to it in the real world.
The QFI associated with the estimation of  parameters, such as (linear) loss or absorption, have such a linear dependence on $N$ to start with.

The root of the linear scaling  in Eq.~\eqref{eq:lin} lies in the geometry of the space of CPTP maps $\Lambda_{\mu}$~\cite{Demkowicz-Dobrzanski2012}.
This space is convex, that is, if  $\Lambda_{\mu},\Lambda'_{\mu}$ are CPTP maps, then so is 
$\Lambda^{\mathbb{p}}_{\mu} = \mathbb{p} \Lambda_{\mu} + (1- \mathbb{p}) \Lambda'_{\mu}$
for a probability $\mathbb{p}.$ Furthermore, $\Lambda^{\mathbb{p}}_{\mu} $ can be realised physically by 
randomly applying $\Lambda_{\mu}$ or $\Lambda'_{\mu}$ with probabilities $\mathbb{p}$ and  $1-\mathbb{p}$ respectively.
Thus, the CPTP map corresponding to a quantum sensor subject to noise or decoherence acting on a $N$-photon quantum state of light
is equivalent to a convex combination of $N$ separate maps acting on single photons.
In particular, these separate maps are independent of $\mu$ which only enters through the  probabilities describing the convex combination.
This essentially makes the sensing task one of classical estimation, and leads to the linear scaling in Eq.~\eqref{eq:lin}.

Maps that cannot be decomposed into a convex combination of other maps are called extremal. 
An instance of an extremal map is a unitary such as $U_{\phi} $ which lead to the quadratic scaling in Eq.~\eqref{eq:sq}.
The linear scaling in Eq.~\eqref{eq:lin} is thus, if possibly abstract, of most fundamental origins and seemingly insurmountable.
Consequently, all stakeholders in the field of quantum sensing must recognise it.

The constant $c$ has been evaluated analytically for widely prevalent processes such as photon loss, dephasing, spontaneous emission~\cite{Demkowicz-Dobrzanski2012}. 
These suggest that diminishing losses and decoherence increase the value of $c.$ 
While its exact behaviour depends on the details of the sensor as well as the loss and noise processes, the magnitude of $c$ can be increased with improved technology, 
leading to possibly substantial tangible quantum enhancements in sensing.

\section{Sensing modalities}
\label{sec:mod}

In my view, the future of quantum-enhanced sensing rests on increasing the constant $c.$
This depends on the specifics of the sensing modality.
From the domain of umpteen sensing modalities~\cite{Degen2017,Yu2021}, even with light~\cite{Taylor2016,Lawrie2019,Berchera2019,Lee2021,Kim2023,Defienne2024}, 
I will focus on three in this section. 
If this suggests that the drive for quantum-enhanced sensing to be platform and application specific, it is only partially true. 
In the following, I will highlight some general concepts and insights that are applicable across sensing domains 
and modalities while restricting myself to three common paradigms with relevance for nanophotonics.

\subsection{Interferometry}
\label{subsubsec:ifo}

\begin{figure}
\includegraphics[width=9cm]{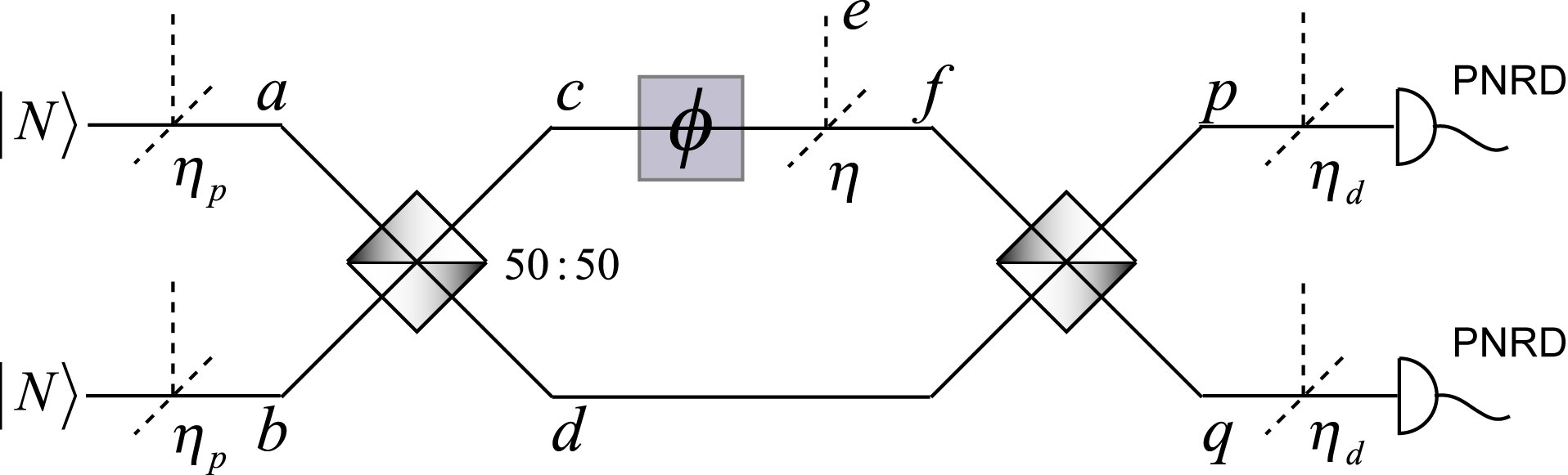}
\caption{Relative phase $\phi$ estimation in a linear interferometer, where $\eta_p,\eta,\eta_d$ are the preparation, transmission, and detection losses respectively. $\eta_p=\eta=\eta_d=1$ denotes no losses. Taken from Ref.~\cite{Datta2011}. }
\label{fig:setup}
\end{figure}

\begin{figure}[h!]
\includegraphics[width=4.25cm]{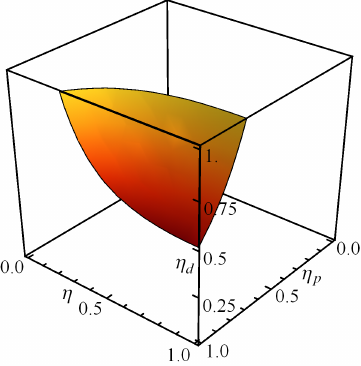}
\includegraphics[width=4.25cm]{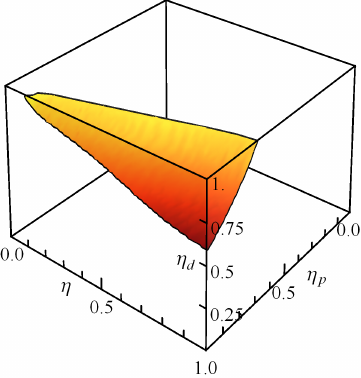}
\caption{Region of $(\eta_p,\eta,\eta_d)$ space of losses for which quantum enhancement is possible using $N=1$ (left) and $N=3$ (right) in the sensor in Fig.~ \ref{fig:setup}.
 Taken from Ref.~\cite{Datta2011}.}
\label{fig:hb3}
\end{figure}

I begin with the estimation of a relative phase, the natural sensing task emanating from interferometry.
A real linear interferometer, especially implemented using integrated photonics is subject to losses~\cite{ThomasPeter2011,Datta2011}. 
These are typically assumed to be linear, and associated with the preparation $\eta_p$ and detection $\eta_d$, as well as the sensing of the phase $\phi$ itself 
$\eta.$ The performance of the quantum sensor in Fig.~\ref{fig:setup}, with fixed-photon-number state inputs and photon-number-resolving detectors (PNRDs)
is quantified by the CFI of estimating $\phi$ using the formalism of Sec.~\ref{sec:qest}.
For quantum enhancement, this must exceed the CFI of estimating $\phi$ using classical light with $|\alpha|^2 = 2N$, as noted in Fig.~\ref{fig:layout}.
This is only possible if the losses are such as to lie within a region of the unit cube of $\eta_p,\eta,\eta_d$ as shown for $N=1$ and $N=3$ in Fig.~\ref{fig:hb3}.
I extract two observations from them.

The first observation is that the most demanding call is placed on the detector efficiency $\eta_d.$
In other words, high-efficiency PNRDs are necessary for phase estimation with quantum light using a linear interferometer as envisaged in Fig.~\eqref{fig:setup}, to outperform its classical counterpart.
This is because the most damage to a quantum state in sensor occurs when the quantum light has picked up all the information it can about the parameter to be estimated.

Integrated, high-efficiency PNRDs have become available in recent years~\cite{Hpker2019}.
To minimise losses in the preparation, the source of the quantum light can be placed on the same integrated chip.
This is also becoming possible~\cite{Stokowski2023}. 
Finally, placing the sample on the same platform is essential to fulfil the vision of quantum nanophotonic sensing, especially in biomedical settings.
This remains a challenge even in nanophotonic biosensing with classical light~\cite[Fig. 4]{Altug2022} as is that of having sources and detectors operating at light wavelengths 
most relevant for the sample~\cite{Todorov2024}.

The second observation is that the demand on $\eta_d$ is greater for the larger $N.$ The same applies to $\eta_p.$
While having more photons does allow greater protection from transmission losses $\eta,$ the performance of the sensor is limited by its most demanding component $\eta_d.$
Quantum states with more photons are by themselves thus not enough for a better quantum sensor.
Indeed, classical states of light are more robust against losses than quantum states.
Better quantum sensors thus need detectors with efficiency increasing with $N$, and number resolvability. 
These do not make for a scalable approach for quantum enhanced sensing.

Quantum error correction might counter losses in a scalable manner. 
Some theoretical works have shown that quantum error correction can indeed combat certain kinds of noise processes in sensing~\cite{Preskill2000,Dur2014,Arrad2014,Herrera2015}. These and other works~\cite{Zhou2018} find that the elusive quadratic scaling of Eq.~\eqref{eq:sq} 
can be recovered, but require unphysical error correcting operations such as infinitely fast ones.
They also rely on unphysical assumptions such as perfect error correcting operations.
Experimentally, quantum error correction does help improve quantum sensing for small $N$ by combating certain kinds of noise in nitrogen-vacancy centre based magnetometry~\cite{Unden2016}.
Similar demonstrations against loss in sensing with quantum light would be exciting.

All real-world error-correction operations are themselves error prone.
Combating that requires fault tolerance. Fault-tolerant quantum sensing has been explored theoretically~\cite{Kapourniotis2019}.
It identified two categories of noise: (i) beyond our control, associated with sensing the parameter, and 
(ii) under our control, associated with in operations such as preparing, manipulating, and measuring probes and ancillae.
It then introduced noise thresholds to quantify the noise resilience of parameter estimation schemes, and theoretically demonstrated
improved noise thresholds over the non-fault-tolerant schemes. 
More importantly, It showed that better devices, which can be engineered under our control, 
can counter larger noise beyond our control.
The use of bespoke quantum error correcting codes for combating losses would identify the efficacy of fault-tolerant sensing with quantum light.

Despite the specificity of the sensing modality in this subsection, I hope that both the importance of detectors and centrality of combating
losses in a scalable manner are general aspects of sensing with quantum light worth recognising.
Quantum sensor designers may feel the suggested remedy of deploying the machinery of quantum fault tolerance excessive and foreign.
This feeling is perhaps nurtured by the belief that building a quantum sensor ought to be an easier endeavour than building a quantum computer, 
the domain to which fault tolerance is typically associated.
As quantum light with more photons become available and losses which are determined by material properties of sensors and detectors cannot be reduced in proportion, 
experimenters may be forced to tread the path traced above.

\subsection{Microscopy}
\label{sec:micro}

I now move to microscopy with quantum light.
The set of possibilities is immense even within optical microscopy.\footnote{I will not address quantum-inspired spatial superresolution of optical sources~\cite{Tsang2016,Tsang2017} as it does not rely on the use of quantum light but rather on spatial mode-resolved measurements.
These mode-resolved measurements are mathematically akin to their temporal or spectral counterparts noted in Sec.~\ref{spectro}.}
Following the strategy noted in Fig.~\ref{fig:layout}, I focus on microscopy with quantum light that improves upon classical schemes limited by shot noise~\cite{Min2009,Freudiger2014}.
Shot noise is attributed to the Poissionian photon number distribution of the coherent state and identified with the linear scaling of the QFI for coherent states  noted in Sec.~\ref{sec:elus}.
Brigher coherent state pulses can give more precision but lead to greater photodamage of the sample, a particular concern in biomedical applications.
The strength of the coherent state thus limited, the purpose of microscopy with quantum light is to continue improving the precision.

Quantum enhanced stimulated emission microscopy~\cite{Triginer_Garces2020} 
and quantum enhanced stimulated Raman spectroscopy~\cite{deAndrade2020,Casacio2021}
achieve this goal. 
The latter work also undertook a detailed study of photodamage. 
These detect the concentration of a sample, which is proportional to the number of photons absorbed (or lost) from a probe pulse.
Mathematically, the best quantum state for estimating loss is a Fock state with a fixed number of photons, say $\ket{n}$~\cite{Adesso2009}.
This because on the quantum state after the loss is $\hat{a}\ket{n} \propto \ket{n-1},$ which is orthogonal to $\ket{n}$ and
 $\hat a$ is the annihilation operator of the appropriate field mode.
On the contrary, the coherent state is the worst, because $\hat{a}\ket{\alpha} \propto \ket{\alpha},$ making it less sensitive to the loss.
The microscopy experiments~\cite{Triginer_Garces2020,Casacio2021} use pulses of intensity-squeezed coherent states of light.
This is a judicious compromise between the mathematically optimal and physically practical -- an abiding principle to be followed in any real world undertaking. 

Another abiding principle, essential for sensing with quantum light to be worthwhile, is to operate in a shot-noise-limited regime. 
In other words, the classical method must be limited in a manner that quantum ones can improve upon. 
This is typically not the case.
Indeed, operating a shot-noise-limited classical sensor may be harder in practice than imposing quantum enhancements on it.
In many optical sensors as in the microscopy schemes noted above, the classical noise sources such as technical laser noise 
and photodetector electronic noise dominate the shot noise at low frequencies.
This is overcome by modulating the signal at a few megaHertz where shot noise is dominant.
Intensity-squeezed coherent states of light then perform quantum-enhanced microscopy.

Generating intensity-squeezed coherent states of light is easier than high-$n$ Fock states, but harder than coherent states.
The challenge arises due to incomplete understanding of nonlinear process of a pulsed nature such as 
group velocity mismatch between the pump and seed pulses, beam divergence, spatial walk-off, or spatiotemporal coupling in the parametric gain~\cite{Triginer_Garces2020}.
Furthermore, higher gain, which is necessary for more squeezing and hence quantum enhancement, requires a more thorough analysis~\cite{Quesada2020}
than low gain efforts providing a squeezing of a dB or lower.
This ties back to the challenge of producing quantum states of light with large photon numbers and high repetition rates, as noted in  Sec.~\ref{sec:elus}.

This modality once again highlights the challenge of generating quantum states of light with large numbers of photons.
Another crucial issue in sensing with quantum light is identifying the limitation of the corresponding sensor using classical light.
It must be that the classical nature of the light -- shot noise is the limitation in the classical sensor.
Attaining that limit, which may be a nontrivial experimental and technological step in itself, is essential before embarking on sensing with quantum light.

\subsection{Spectroscopy}
\label{spectro}

I finally turn to spectroscopy, perhaps the most ubiquitous of analytical tools in the natural sciences. 
I also expand to a paradigm where $\Lambda_{\mu}$ captures the interaction between quantum light and quantum matter from the first principles of quantum mechanics.
Effective quantities such as phase and loss discussed the last two sensing modalities
are functions of the underlying `microscopic' parameters of the quantum matter system such as transition energies, line widths or lifetimes, 
dipole-dipole couplings, electron-phonon couplings, etc.
This expanded paradigm leads to an operational definition of spectroscopy as the precise estimation of these `microscopic' parameters.
It is also more informative when novel complex quantum matter systems are studied spectroscopically, than merely estimating effective macroscopic parameters such as absorption and refractive index.

Contemporary theoretical studies of spectroscopy treat the light classically, leading to a semiclassical light-matter interaction~\cite{Mukamel1999}.
Formally, the total hamiltonian
\be
\label{eq:ham}
\hat H = \hat H^{\mathrm{M}}_{\mu} + \hat H^{\mathrm{L}} + \hat H^{\mathrm{LM}},
\ee
captures the operation $\Lambda_{\mu}[\rho_{\mathrm{cl}}],$
where $\hat H^{\mathrm{M}}_{\mu}$ is the hamiltonian of a matter system or sample with parameters $\mu$, 
and $\hat H^{\mathrm{L}},\hat H^{\mathrm{LM}}$ are those of the light and light-matter interaction respectively.
For a chosen measurement $\bm{\Pi}$ corresponding to a spectroscopic configuration, the theoretically evaluated signal $\Tr\left[\Lambda_{\mu}[\rho_{\mathrm{cl}}]\bm{\Pi} \right]$ is compared with experimental data to estimate
$\mu$ by data fitting.

Early experiments efforts towards quantum light spectroscopy used classical laser light
 to measure semiconductor quantum wells and processed the data to mimic the response of the system to a quantum state of light~\cite{Kira2011}.
Early theoretical works did study the effect of shining quantum light directly on complex quantum systems.
They evaluated the signal for specific spectroscopic configurations and measurement schemes~\cite{Dorfman2016},
and showed that some excitation transfer pathways in coupled quantum systems obscured in nonlinear spectroscopy using  classical light could be revealed by using quantum light.
As these studies work with specific input states and measurements, they are incapable of optimising over input states and measurements as outlined in Sec.~\ref{sec:qest} and Fig.~\eqref{fig:layout}.

\begin{figure}[h!]
\includegraphics[width=8.0cm]{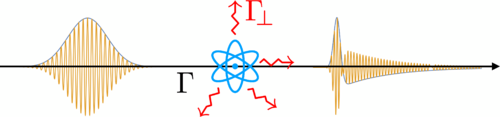}
\caption{ Illustration (not to scale) of the excitation of quantum matter system by a quantum pulse of light (with Gaussian temporal envelope). 
$\Gamma$ represents the interaction strength with the pulse, while $\Gamma_{\perp}$ describes emission into other (inaccessible) orthogonal field modes. 
As illustrated, the shape of the wave packet is changed by the interaction with the matter system. Taken from Ref.~\cite{Albarelli2023}}
\label{fig:spec}
\end{figure}

Recent works have cast spectroscopy in the framework of quantum estimation theory to begin traversing the path outlined in Fig.~\eqref{fig:layout}.
It relies on calculating the quantum state of the light pulse $\Lambda_{\mu}[\rho]$ after it has interacted with a quantum matter system, as in Fig.~\eqref{fig:spec}.
This the central ingredient in the evaluation of the QFI of estimating `microscopic' parameters of the matter system and identification of optimal measurements~\cite{Albarelli2023,Darsheshdar2024}.
These works reproduce known results on absorption spectroscopy when a two-level system (TLS) starts in the ground state, and should do so for stimulated emission spectroscopy when the TLS starts in an excited state. 

The new insight from this formalism is that additional information about the matter system is available in the distorted quantum light pulse after the interaction.
This is in addition to the information that can be extracted from absorption or emission spectroscopy. In other words, beyond counting the number of photons, 
there is more information in the shape of the pulse itself that can be extracted from quantum light spectroscopy, as illustrated in Fig.~\eqref{fig:spec}.
The relative contributions depends on the time duration of the experiment relative to the lifetime, as well as the fraction  $\Gamma/\Gamma_{\perp}$  of the light detected.

Another novel insight is that higher precision in estimating a parameter such as lifetime is not coincident with higher excitation probability or cross section~\cite{Albarelli2023}.
This is significant as the latter is often taken to be a surrogate for the efficacy of spectroscopy~\cite{Raymer2021}.
Other insights can be extracted using entangled quantum states as probes such as biphotons and more involved matter systems such as coupled dimers~\cite{Khan2024}.

\begin{figure}
\includegraphics[width=8.0cm]{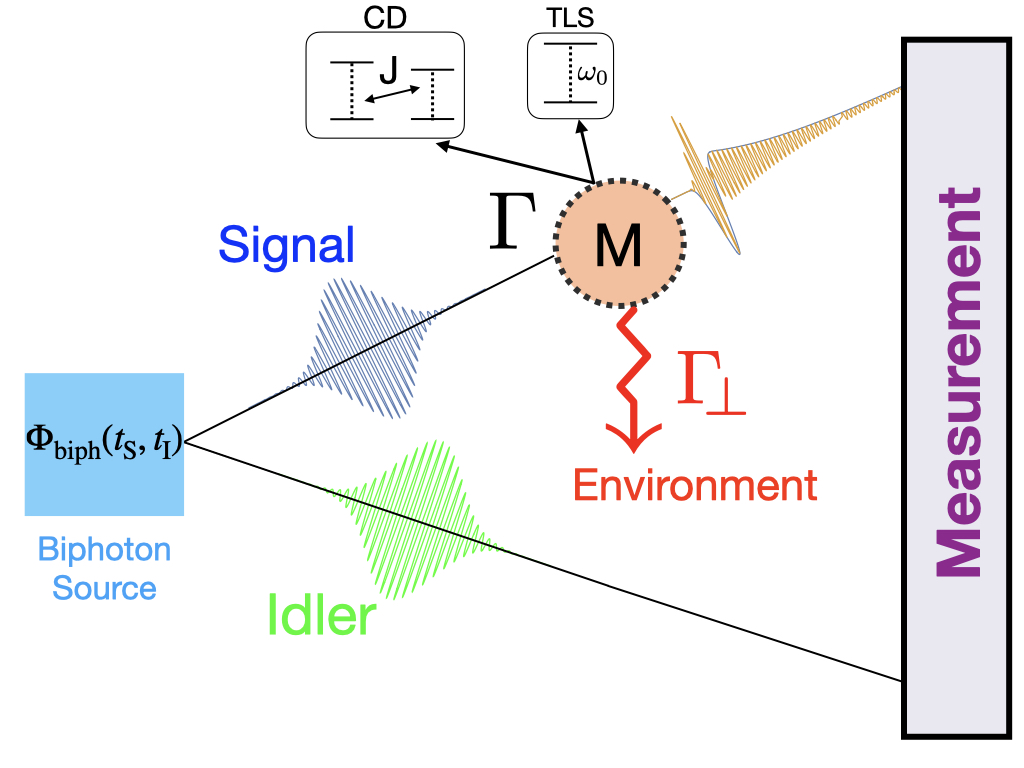}
\caption{Schematic of pulsed biphoton spectroscopy. Taken from Ref.~\cite{Khan2024}}
\label{fig:biph}
\end{figure}

Extracting this additional information from the pulse shape requires temporal or spectral mode-resolved measurements.
Such measurements are now available~\cite{Serino2023} for single photons.
Non-destructive photon number counting is necessary to  extract the information from absorption or emission spectroscopy simultaneously.
This is more challenging for single photons at optical wavelengths.
The strength of the formalism also lies in showing what is unnecessary. For instance, entangled measurements are not necessary to attain the best fundamental precision if only one mode of a biphoton state interacts with the sample, as in Fig.~\eqref{fig:biph}.

To obtain the quantum state $\Lambda_{\mu}[\rho]$ of the pulse of light after the interaction, the dynamical equations of quantum mechanics 
such as the Schr\"odinger equation for the hamiltonian in Eq.~\eqref{eq:ham} must be solved. 
For pulsed interactions, $\hat H^{\mathrm{LM}}$ is time-dependent while for complex quantum systems,
$\hat H^{\mathrm{M}}_{\mu}$ typically consists of electronic and phononic contributions as well as their couplings.
Accounting for the latter requires the use of hierarchical equations of motion~\cite{Tanimura1990}.
Coupling this with incident quantum states of light with indefinite photon numbers such as squeezed states requires tools such as tensor networks to express
$\Lambda_{\mu}[\rho]$ efficiently such that their QFI can be evaluated~\cite{Khan2023b}.
This moves us into realm of quantum light-matter interactions~\cite{GonzlezTudela2024}, a rich field of study in itself.

Quantum light spectroscopy is deeply intertwined the study of quantum light-matter interactions, especially in quantum nanophotonics. 
The latter engineers configutaions that can enhance the light-matter interaction to probe the physics of quantum matter systems such as complex molecules.
Cavities, waveguides, and surfaces provide the varied platforms to perform quantum light spectroscopy, including providing sources of quantum light.
Combining these with integrated photonic setups with efficient detectors~\cite[Fig. 10]{Chang2023} could vastly expand the arena of quantum light spectroscopy into domains such as 
quantum nanophotonic biosensing~\cite{Altug2022}.


\section{Conclusions}
 \label{sec:Conc}

I conclude with an enumerated gist of the vital challenges and possible avenues for overcoming them.
\begin{enumerate}

\item Detecting quantum states of light with high efficiency. This is to make the most of the available quantum states of light. 
The efforts will have to be in material science and technology.

\item Generating `bright' quantum states of light with large numbers of photons at repetition rates and determinacy comparable to classical sources. 
The efforts should be in developing sources of `bright' squeezed states of light.

\item Combat inevitable losses in the sensing process by develop error correction and fault tolerance for quantum sensing. 
The efforts will have to be in quantum information and coding theory. 

\item Integrating sources, interactions, and detectors on the same platform. This will be pulled by specific applications and pushed by fabrication technologies.

\end{enumerate}

\begin{acknowledgments}
I thank my numerous colleagues and collaborators for stimulating discussions over the years.
This work was supported, in part, by grant from the UKRI (Reference Number: 10038209) under the UK Government’s Horizon Europe Guarantee for the European Union's Horizon Europe Research and Innovation Programme under agreement 101070700 (MIRAQLS).
\end{acknowledgments}

\bibliographystyle{apsrev4-2}
\bibliography{Refs}

	\end{document}